\documentstyle[12pt]{article} 
\begin{document} \newcommand {\ber}
{\begin{eqnarray*}} \newcommand {\eer} {\end{eqnarray*}}
\newcommand {\bea}
{\begin{eqnarray}} \newcommand {\eea} {\end{eqnarray}} \newcommand {\beq}
{\begin{equation}} \newcommand {\eeq} {\end{equation}}
\newcommand {\state}
[1] {\mid \! \! {#1} \rangle} \newcommand {\eqref} [1] {(\ref {#1})}
\newcommand{\preprint}[1]{\begin{table}[t] 
           \begin{flushright}               
           \begin{large}{#1}\end{large}     
           \end{flushright}                 
           \end{table}}                     
\def\Acknowledgements{\bigskip  \bigskip {\begin{center} \begin{large}
             \bf ACKNOWLEDGEMENTS \end{large}\end{center}}}

\newcommand{\half} {{1\over {\sqrt2}}}
\def\Dslash{\not{\hbox{\kern-4pt $D$}}}
\def\nc{N_c}
\def\nf{N_f}
\def\tbc{{\bf To Be Completed}}
\def\CR{\nonumber \\ }
\def\Q{$QCD_2\ $}
\def\cmp#1{{\it Comm. Math. Phys.} {\bf #1}}
\def\cqg#1{{\it Class. Quantum Grav.} {\bf #1}}
\def\pl#1{{\it Phys. Lett.} {\bf #1B}}
\def\prl#1{{\it Phys. Rev. Lett.} {\bf #1}}
\def\prd#1{{\it Phys. Rev.} {\bf D#1}}
\def\prr#1{{\it Phys. Rev.} {\bf #1}}
\def\np#1{{\it Nucl. Phys.} {\bf B#1}}
\def\ncim#1{{\it Nuovo Cimento} {\bf #1}}
\def\jmath#1{{\it J. Math. Phys.} {\bf #1}}
\def\mpl#1{{\it Mod. Phys. Lett.}{\bf A#1}}
\def\jmp#1{{\it J. Mod. Phys.}{\bf A#1}}
\def\mycomm#1{\hfill\break{\tt #1}\hfill\break}

\begin{titlepage}
\titlepage
\rightline{TAUP-2285-95}
\rightline{\today}
\vskip 1cm
\centerline{{\Large \bf Mesonic  Spectra of Bosonized $QCD_2$ Models}}
\vskip 1cm

\centerline{A. Armoni and J. Sonnenschein
\footnote{Work supported in part by the Israel Academy of Science,
and  the US-Israel Binational
Science Foundation }}

\vskip 1cm
\begin{center}
\em School of Physics and Astronomy
\\Beverly and Raymond Sackler Faculty of Exact Sciences
\\Tel Aviv University, Ramat Aviv, 69978, Israel
\end{center}
\vskip 1cm
\begin{abstract}
 We discuss bosonized  two-dimensional
 QCD with massless fermions
 in the adjoint and  multi-flavor
fundamental representations. We evaluate the
 massive mesonic  spectra of several models by using the  light-front
quantization and
diagonalizing   the mass
operator $M^2=2P^+P^- $.
 We recover previous results in the case of one flavor adjoint
fermions
 and we find the exact massive spectrum of multi flavor QCD in the limit
of large  number of flavors.
\end{abstract}
\end{titlepage}
\newpage

\section{Introduction}

QCD in two space-time dimensions ($QCD_2$)  is believed to be a useful
laboratory
to  examine ideas about the hadronic physics of the real world.
One such an idea,  the  expansion in  large number of colors $N_c$,
was applied in the pioneering work of
  't Hooft \cite{thooft}.
A mesonic spectrum  of a ``Regge trajectory"  type was discovered in that work
for  $QCD_2$  with Dirac fermions in the
 fundamental representation.
The large $N_c$ limit is taken while keeping $e^2N_c$  fixed, where $e$ is the
 gauge coupling.  The ``orthogonal" approach,  the strong coupling
limit, was also found to be a fruitful technique
when combined with bosonizing the system.
In that framework the low energy effective
action of the theory can be  derived  exactly.  Quantization
of soliton solutions of  that  effective theory led to the low lying
semi-classical baryonic spectrum\cite{DFS}.

The large $N_c$ approach combined with a light-front quantization
was applied to the study of \Q with Majorana fermions in the adjoint
representation.  Unlike the case of fundamental fermions, for adjoint fermions
pair creation is not suppressed in the large $N_c$ limit. The bosonic spectrum
includes  an infinite number of approximately linear Regge trajectories which
are associated with an exponential growing density of states at
high-energy\cite{kutasov,gyan}. Recently, a ``universality" behavior of \Q was
found in the sense that the physics     of massive  states depends only on the
gauge group and  the afine Lie algebra level but not on the
representation of the group\cite{kutasov2}.

In the present paper we combine the bosonization technique with that of a
 large $N$ expansion and a light-front quantization
in the analysis of the massive mesonic spectrum  of several \Q models.
The massless sector  which was discussed in ref.\cite{kutasov2}
is not addressed  in the present work.
The models include  massless  fermions in the  adjoint representation
and  multi flavor fundamental representations.
In the former case we expand in  $\nc-$ the
 number of colors whereas in the latter case we consider large $\nf-$
 number of flavors .
In case of massless fermions the bosonization formalism is convenient since it
separates the color, flavor and baryon number degrees of freedom\cite{FSrep}.
Moreover the generalization from fermions coupled to YM fields
to an arbitrary  gauged afine Lie algebra system is natural in  the bosonized
picture.
The basic difference between our approach and the one taken in refs.
\cite{kutasov,gyan} is the use of current quanta rather then those of
quarks in constructing  the mass operator $M^2$ and thus also the wave equation
and
the Hilbert space.
Our results are in agreement with the results of refs.\cite{kutasov,gyan} for
the case of adjoint fermions. In the large $N_f$ it is shown that
the exact massive spectrum is a single particle with
$M^2={{e^2 N_f} \over \pi}$.
This phenomenon is explained by the fact that this limit can be viewed as
an ``abelianization" of the model.

 The  paper is organized as follows. In section 2 we present
the models. The actions are written down in their bosonized versions.
  We then derive
 explicit expressions for the momentum operators and the mass operator.
The afine Lie algebra currents are expanded in terms of annihilation
and creation
 operators with which  a Fock space of physical states is built.
 In section 3 the model with adjoint fermions is analyzed in the
large $\nc$ limit.
We introduce wave functions  and write down an eigenvalue equation which
generalizes 't Hooft equation\cite{thooft}. The massive mesonic spectrum
was then deduced.
 In section 4 we find that the  exact spectrum  of multi-flavor QCD
with fundamental fermions in the regime of  $N_f \gg N_c$ include only one
single mesonic state. We discuss the relation of this spectrum
to that of  multi-flavor QED.
The case of large level $WZW$ models  is also considered.
Some conclusions and certain open problems are discussed in section 5.

\section{The Models}
Consider two dimensional $SU(N_c)$ Yang-Mills gauge fields
 coupled to (i) $N_f$ massless Dirac
fermions in the fundamental representation, or (ii) massless Majorana
 fermions in the adjoint representation.
These theories are described by the following  classical Lagrangian:
\beq
{\cal L}=-{1\over {4e^2}}Tr[ F_{\mu\nu}^2+i\bar \Psi\Dslash\Psi]
\eeq
where $F_{\mu\nu}=\partial _\mu A_\nu - \partial _\nu A_\mu + i[A_\mu ,A_\nu]$
and the trace is  over the
color and flavor indices.  For the   case (i)
$\Psi$ has the following  group structure $\Psi_{ia}$ where $i=1,...,\nc$
and $a=1,...,\nf$, $D_\mu=\partial_\mu -iA_\mu$
 whereas  for
the  case  (ii)  $\Psi\equiv \Psi^i_j$
and $D_\mu=\partial_\mu -i[A_\mu,\ ]$.
 In both cases
 $\Psi$ is a two spinor parametrized as follows
$\Psi = \left( \begin{array}{c} \bar \psi \\ \psi \end{array} \right) $.

It is useful to handle these models in the framework of
the light-front quantization, namely, to use light-cone space-time
  coordinates and to choose  the chiral gauge
 $A_-=0$. In this scheme the quantum
 Lagrangian takes the form
\beq
{\cal L}=-{1\over {2e^2}} {(\partial _-A_+)}^2+i\psi ^\dagger \partial_+\psi
+i\bar \psi ^\dagger \partial_-\bar \psi + A_+J^+
\eeq
where color and flavor indices were omitted and $J^+$ denotes the $+$
component of the color current
 $J^+ \equiv \psi ^\dagger \psi$.

By choosing $x^+$ to be the `time' coordinate it is clear that $A_+$
and $\bar \psi$ are non-dynamical degrees of freedom.
In fact, $\bar \psi$ are  decoupled from the other fields
so in order to
 extract  the physics of the dynamical degree of freedom
 one has to functionally integrate  over
$A_+$. The result of this integration is  the following simplified Lagrangian
\beq
{\cal L}={\cal L}_0+{\cal L}_I=i\psi ^\dagger \partial_+\psi+i\bar\psi ^\dagger
\partial_-\bar\psi
 -e^2J^+{1\over {\partial _- ^2}}J^+\label{SimAc}
\eeq

Since our basic idea is to solve the system in terms of  the ``quanta" of the
colored  currents, it is natural to introduce bosonization descriptions of the
various models.

(i) The bosonized action  of colored flavored Dirac fermions in the fundamental
representation is expressed in terms of a WZW\cite{witten} action of a group
element  $u\in U(\nc\times\nf)$ with an additional mass term that couples the
color, flavor and baryon number sectors\cite{FS2}. In the massless case when
the latter term is missing the action takes the following form
\beq
S^{fun}_0=S^{WZW}_{(\nf)}(g)  + S^{WZW}_{(\nc)}(h)+
{1 \over 2} \int d^2x \partial _\mu \phi \partial ^\mu \phi\label{BLfc}
\eeq
where $g\in SU(\nc),\  h\in SU(\nf)$ and $e^{i\sqrt{4\pi\over \nc\nf}\phi}\in
U_B(1)$ with $U_B(1)$ denoting the baryon number symmetry and
\ber
\lefteqn{S^{WZW}_{(k)}(g)={k\over{8\pi}}\int _\Sigma d^2x \ tr(\partial _\mu
g\partial ^\mu g^{-1}) + } \\
 && {k\over{12\pi}}\int _B d^3y \epsilon ^{ijk} \
 tr(g^{-1}\partial _i g) (g^{-1}\partial _j g)(g^{-1}\partial _k g),
\eer

(ii)  The current structure of free Majorana fermions in the adjoint
representation can be recast in terms of a WZW action of level $k=\nc$,
 namely, $ S^{adj}_0=S^{WZW}_{(\nc)}(g)$ where now $g$ is in the adjoint
representation of $SU(N_c)$, so that it carries a conformal dimensions of
 ${1\over 2}$\cite{usss}.

Multi-flavor adjoint fermions can be described as $S^{WZW}_{N_f}(g) +
S^{WZW}_{N_c^2-1}(h)$ where $g\in SO(N_c^2-1)$ and $h\in SO(N_f)$. In the
present work we discuss only gauging of $SU(N_c)$ WZW so the latter model
would not be considered.

Substituting now $S^{fun}_0$ or  $S^{adj}_0$ for $S_0$ the action that
corresponds to \eqref{SimAc} is given by
\beq
S= S_0-{e^2 \over 2}\int d^2x J^+{1\over{\partial _- ^2}}J^+
\eeq
where
 the current
$J^+$ now reads $J^+=i{k\over{2\pi}}g\partial _- g^\dagger$,
 where the level $k=\nf$ and  $k=\nc$
for the multi-flavor fundamental and adjoint cases respectively.

The light-front quantization scheme is very convenient because the
corresponding momenta generators $P^+$ and $P^-$ can be expressed only in terms
of $J^+$. We would like to emphasize that this holds only for  the massless
case.

Using the Sugawara construction, the contribution of the colored currents to
the momentum operator, $P^+$,
 takes the following  simple form:
\beq
P^+={1\over{N+k}}\int dx^- :J^i_j(x^-)J^j_i(x^-):
\eeq
where  $J \equiv \sqrt \pi J^+$,
 $\nc$
in the denominator is the second Casimir operator of the adjoint
representation and  the level $k$ takes the values mentioned above.
Note that for future purposes we  have added the  color  indices $i,j=1
\ldots \nc$ to the currents.
In the absence of the interaction with the gauge fields
the second momentum operator, $P^-$ vanishes. For the  various
$QCD_2$ models it is given by
\beq
P^-=-{e^2 \over {2\pi}}\int dx^-:J^i_j(x^-){1\over{\partial _- ^2}}J^j_i(x^-):
\eeq

In order to find the massive
spectrum of the model we should diagonalize the mass
operator $M^2=2P^+P^-$.
 Our task is therefore  to  solve the eigenvalue equation
\beq
2P^+P^-\state{\psi} = M^2\state{\psi}.
\eeq
We write $P^+$ and $P^-$ in term of the Fourier transform of $J(x^-)$
defined by $J(p^+)=\int {dx^- \over {\sqrt{2\pi}}} e^{-ip^+x^-} J(x^-)$.
Normal ordering in the expression of $P^+$ and $P^-$ are naturally with respect
to $p$, where $p<0$ denotes a creation operator.
 To simplify the notation we
write from here on $p$ instead of $p^+$. In terms of these variables the
momenta
generators are   \bea
&P^+={2\over{N+k}}\int ^\infty _0dp J^i_j(-p)J^j_i(p)     \CR
&P^-={e^2 \over \pi}\int ^\infty _0dp {1\over{p^2}}J^i_j(-p)J^j_i(p)
\label{P}
\eea
Recall that the light-cone currents $J^i_j(p)$ obey a level $k$, $SU(\nc)$
affine Lie algebra \beq
[J^k_i(p),J^n_l(p')]={1\over 2}kp(\delta ^n_i\delta ^k_l - {1\over N}\delta
^k_i
\delta ^n_l)\delta (p+p')+
\half (J^n_i(p+p')\delta ^k_l - J^k_l(p+p')\delta ^n_i)
\label{Kac}
\eeq
We  can now construct the Hilbert space. The vacuum $\state{0,R}$ is defined
by the annihilation property:
\beq
\forall p>0,\ J(p)\state{0,R}=0
\eeq
Where R is an ``allowed"  representations depending on the level\cite{gepner}.
Thus, a typical state in Hilbert space is \\
$Tr\ J(-p_1)\ldots J(-p_n)\state{0,R}$.

Diagonalizing $M^2$, is in general, a complicated task, hence we will
 examine in detail some special cases of the theory.
\section{Large $N_c$ Adjoint Fermions}
The  first case we analyze is $QCD_2$ with fermions in the adjoint
representation. This model was investigated in the past \cite{kutasov}
\cite{gyan} and recently it was shown \cite{kutasov2}
that its massive spectrum is the same as the model of $N_f=N_c$ fundamental
 quarks.

Since an exact solution of the model is beyond our reach we employ  a large
$N=N_c$ approximation where some simplifications occur. As stated above the
$N_c=N$ is also the corresponding level of the model.
First, we write down the general symmetric singlet states:
\ber
\state{\psi} = \sum ^\infty _{n=2} {1\over {N^n}} \int ^P _0 \ldots \int ^P _0
dp_1 \ldots dp_n \delta (P-\sum ^n _{i=1} p_i) \phi _n (p_1,\ldots ,p_n)\times
\\ \times {1\over n} \sum_\sigma J^{j_1}_{j_2}(-p_{\sigma (1)})
J^{j_2}_{j_3}(-p_{\sigma(2)}) \ldots J^{j_n}_{j_1}(-p_{\sigma (n)}) \state{0}
\eer

\ber
\state{\psi} '= \sum ^\infty _{n=2} {1\over {N^n}} \int ^P _0 \ldots \int ^P _0
dp_1 \ldots dp_n \delta (P-\sum ^n _{i=1} p_i) \chi _n (p_1,\ldots ,p_n)\times
\\ \times {1\over n} \sum_\sigma J^{j_1}_{j_2}(-p_{\sigma (1)})
J^{j_2}_{j_3}(-p_{\sigma(2)}) \ldots J^{j_n}_{j_{n+1}}(-p_{\sigma (n)})
\state{0} ^{j_{n+1}}_{j_1}
\eer

Here $\phi _n,\chi _n$ are the Schr\"{o}dinger wave functions, and the
summation over $\sigma$ is, as explained below, over the cyclic
group $Z_n$. The difference
between the two states is that in $\state{\psi}$ the currents act on
 the identity vacuum,
 whereas in $\state{\psi} '$ the currents act on the adjoint vacuum.
The latter is
a state that transforms in the adjoint representation of $SU(N)$ and obeys
$J(p) \state{0} ^i_j=0$ for all $p>0$.
The two states form two distinct equivalence classes, in a sense that
one state can
not be related  to another one by acting with an operator which is built from
a finite number  of current operators.
In principle, other vacuum representations are allowed, but is was proven
 \cite{kutasov2}
that in the large $N$ limit it is sufficient to consider only the identity
 vacuum
 and the adjoint vacuum.   Other representations lead to multi particle
 states
which are suppressed in the  large $N$ limit.
Note that $M^2 \state{0}=0$, but $M^2 \state{0} ^i_j = m_0^2 \state{0} ^i_j
\ne 0$.
The reason is that the zero mode of the current $J^i_j(0)$ does not annihilate
the adjoint vacuum and actually $J^k_l(0)J^l_k(0) \state{0} ^i_j = N
\state{0} ^i_j$\cite{gepner} .The factor N is the second Casimir
 of the adjoint representation. Thus $M^2 \state{0} ^i_j = 2P^+P^-
\state{0} ^i_j = 2\times {1 \over {2N}}\times N \times {e^2 \over {2\pi}}
\times N \state{0} ^i_j = {e^2 N \over {2\pi}} \state{0}^i_j $.
This non-zero value of $m_0^2$ will lead to mass splitting between
 eigenvalues of $\state{\psi}$ and $\state{\psi} '$.

The summation of  $\sigma$ over all elements of the cyclic group $Z_n$ was
introduce  to  achieve  the following symmetry property  of the wave functions
\bea
 \phi _n(p_1,p_2,\ldots ,p_{n-1},p_n)=\phi _n(p_n,p_1,p_2,\ldots ,p_{n-1})
  \nonumber \\
 \chi _n(p_1,p_2,\ldots ,p_{n-1},p_n)=\chi _n(p_n,p_1,p_2,\ldots ,p_{n-1})
 \label{bc1}
\eea
Another property of $\phi _n,\chi_n$ is their boundary condition
\bea
 \phi _n(0,p_2,\ldots ,p_n)=0 \nonumber \\
 \chi _n(0,p_2,\ldots ,p_n)=0
 \label{bc2}
\eea
which is  a consequence of hermiticity of the operators\cite{thooft}.
 A general state whether of the type $\state{\psi}$ or   $\state{\psi} '$ is an
eigenstate of $P^+$ with eigenvalue $P$
\beq
 P^+\state{\psi} =P\state{\psi}
\eeq
\beq
 P^+\state{\psi} '=P\state{\psi} '
\eeq
due to the following   commutation
relation
\beq
 [P^+,J^{j_1}_{j_2}(-p)]=pJ^{j_1}_{j_2}(-p).
\eeq
In the more familiar CFT terminology this relation translates into
$[L_0, J_n]= -nJ$.
The calculation of the commutator $[P^-,J^{j_1}_{j_2}(-p)]$ is more
complicated.
In the procedure of evaluating the spectrum we find the eigen wave-function by
solving an integral equation that generally mixes $\phi_n$ (or $\chi_n$) with
different $n$'s, namely, it mixes the number of currents. Since the general
equation is highly non-trivial we will use an approximation in which we
take into account  only
singular terms and terms with the  same  number of current operators.
Those  terms have the  dominant contribution  to  the wave
equation\cite{kutasov}.
A detailed
calculation of the commutators which includes all the terms
is written in the Appendix.
Here we write   only the  most significant  term of the commutator
\bea
\lefteqn {[P^-,J^{j_1}_{j_2}(-p_1)]= } \nonumber \\
&& {e^2\over \pi}\half
\int ^\infty _0 dp({1\over {{(p\!+\!p_1)}^2}} - {1\over p^2})
(J^i_{j_2}(-p\!-\!p_1)J^{j_1}_i(p)-J^{j_1}_i(-p\!-\!p_1)J^i_{j_2}(p))
\nonumber \\
&& + other\ terms
\eea
The above term includes an annihilation operator $J(p)$, which has non-zero
contribution when it commutes with another creation operator, say $J(-p')$.
A non-negligible commutator occurs  only with a `neighbor' of
$J^{j_1}_{j_2}(-p_1)$, namely, a commutator with a  current
which carries   the same group
indices. Other commutators are
 suppressed  by factors of ${1\over N}$. A tedious
though straightforward calculation yields
\bea
&\half \int ^\infty _0 dp({1\over {{(p\!+\!p_1)}^2}} - {1\over
p^2})[(J^i_{j_2}(-p\!-\!p_1)J^{j_1}_i(p)-J^{j_1}_i(-p\!-\!p_1)J^i_{j_2}(p),
J^{j_2}_{j_3}(-p_2)]= \nonumber \\
&{N\over 2}\times \int ^{p_2}_0
dp({1\over {{(p_1\!+\!p_2\!-\!p)}^2}}-{1\over {{(p_2\!-\!p)}^2}})
J^{j_1}_i(-p_1\!-\!p_2\!+\!p)J^i_{j_3}(-p)
+ other\ terms
\eea
were `other terms' are either terms that  are suppressed in large $N$, or terms
that change the number of currents.
We comment on the validity of dropping the latter terms in section 5.
The result of the last two commutators is that
if one starts with two currents $J(-p_1)$ and $J(-p_2)$ he ends with two other
currents which are multiplied by singular denominators. This result leads to
the
following eigenvalue equation
\bea
 M^2\phi _n(p_1,\ldots ,p_n) = -{e^2N \over \pi}\int dy {1\over {(p_1-y)}^2}
 \phi _n(y,p_1\!+\!p_2\!-\!y,p_3,\ldots ,p_n) + \nonumber \\
 cyclic\ permutations \ \ \ \ \ \  \label{thoofteq}
\eea
This  is a generalization of
\begin{em}
  't Hooft's equation.
\end{em}
Obviously, the same equation holds also for $\chi _n$.

Equation \eqref{thoofteq} can be solved analytically  with the boundary
 conditions
\eqref{bc1} and \eqref{bc2}. The simplest solution is for $\phi _2$:
\ber
\phi _2(x) = \sin (\pi kx) & k\in 2Z+1 \nonumber \\
M^2_k = e^2N\pi k
\eer
The general solution for $\phi_{2n}$ is quite involved but the
eigenvalues are rather simple:
\beq
 M^2_{k_1,k_2,\ldots ,k_n}= e^2N\pi (k_1\!+\!k_2\!+\! \ldots \!+\! k_n)
 \ \ \ \  k_1,k_2,\ldots ,k_n \in 2Z+1
\eeq
The result for the $\state{\psi} '$ sector is similar, with the small $m_0^2$
difference:
\beq
 M^2_{k_1,k_2,\ldots ,k_n}= e^2N\pi (k_1\!+\!k_2\!+\! \ldots \!+\! k_n)
 + m_0^2 \ \ \ \  k_1,k_2,\ldots ,k_n \in 2Z+1
\eeq
This expression of the eigenvalues indicates
 an exponential growth of  the number of states, in accordance with previous
results \cite{kutasov}.  There is, however,   a slight difference
between our results and those of ref.\cite{kutasov}. We found that  the
 values of the integers  must be odd  whereas
 in Kutasov's
paper, they are even. The source of this difference is
  the symmetric boundary conditions that we have used \eqref{bc1}.

The spectrum contains two blocks.  The identity-block has integer dimension
and hence can be interpreted as the boson-block,whereas the adjoint-block
 has half integer dimension and thus will be referred as the fermion-block.
We have seen that the two blocks have similar structure
since the
two sectors obey the same wave function equation.
Furthermore, mixture between the two sectors is avoided  due to the
fact that the hamiltonian creates and destroys even numbers of ``quarks".


The``quark" content of the spectrum is determined
by
the relation between  the currents and the quarks. This relation is given by
\beq
 J^i_j(-p)=\int ^\infty _{-\infty } dq \psi ^i_k(q)\psi ^k_j(-p-q)
\eeq
which is the Fourier transform of $J(x)=\psi (x)\psi (x)$, and hence the "boson
-block" can be written as
\ber
\lefteqn{\sum_\sigma J^{j_1}_{j_2}(-p_{\sigma (1)})
J^{j_2}_{j_3}(-p_{\sigma(2)}) \ldots J^{j_n}_{j_1}(-p_{\sigma (n)}) \state{0}
= } \\
&& \sum_\sigma
\int ^\infty _{-\infty} dq_1 \psi ^{j_1}_{i_1}(q_1)
                             \psi ^{i_1}_{j_2}(-p_{\sigma(1)}-q_1)
\int ^\infty _{-\infty} dq_2 \psi ^{j_2}_{i_2}(q_2)
                             \psi ^{i_2}_{j_3}(-p_{\sigma(2)}-q_2)  \\
&& \ldots \int ^\infty _{-\infty} dq_n \psi ^{j_n}_{i_n}(q_n)
                          \psi ^{i_n}_{j_1}(-p_{\sigma(n)}-q_n) \state{0}
\eer
and can be thought  of as a mixture of $2n,2n-2,\ldots ,2$ quarks. In a similar
way, a state in the fermion-block is a mixture of $2n+1,2n-1,\ldots ,3$ quarks.
\section{Large $N_f$ QCD}
Massless $QCD_2$ with
 $N_f$ flavors of quarks in the fundamental representation is
described by the Lagrangian of \eqref{BLfc}. Setting aside the flavor and
baryon number sectors, the left over system is that of a $k=N_f$ level
$SU(\nc)$ WZW action with an additional non-local interaction term.
We are thus led to analyze the spectrum of  the  model with level equal to
$N_f$. In practical terms the latter means substituting $k$ by $N_f$ in
    the expression for $P^+$ eqn.
\eqref{P} and in the Affine Lie algebra eqn. \eqref{Kac}.

The idea is to invoke a large $N_f$ approximation, namely,  to consider models
in which  $k=N_f \gg N_c$.
Models with number of colors and flavors which fall into this regime, are
 significantly simpler than  models that don't obey this inequality.  The basic
reason for that is the simplification of the algebra  eqn. \eqref{Kac}.
 The commutator of $P^-$ with $J$
in the large  $N_f$ limit takes the form
\beq
[P^-,J^i_j(-p)]={e^2 \nf\over {2\pi p}} J^i_j(-p) + e^2 JJ\ term
\eeq
which  means that
\beq
M^2\  {J^i_j(-p) \over {\nf^{1\over 2}}} \state{0,R} =
{e^2\nf \over \pi} \  {J^i_j(-p) \over {\nf^{1\over 2}}} \state{0,R}
+ e^2 \nf^{1\over 2} {JJ \over \nf} \state {0,R}\label{Slnf}
\eeq
Upon neglecting the second term which is suppressed by a factor of ${1\over
{\sqrt\nf}}$   we  get the  following  solution to the  eigenvalue problem
\beq
M^2\  J^i_j(-p) \state{0}^{ja}_{ib} =
                         {e^2\nf \over \pi} \  J^i_j(-p) \state{0}^{ja}_{ib}.
\eeq
Notice that
the current operator
acts
 on the adjoint vacuum and not on the identity vacuum.
The reason for that is obviously the requirement that
states have to be  color singlets.
$\state{0}^{ja}_{ib}$ stands for $\state{0}^{j}_{i}\otimes \state{0}^{a}_{b}$,
namely, the tensor product of the color adjoint vacuum and its flavor
counterpart. Recall that following eqn. \eqref{BLfc} the vacuum state is
 an outer
product of the vacua of the three independent Hilbert-spaces of the color,
flavor and baryon number sectors. This ``decoupling" of these spaces of
states is an artifact of the massless limit of the bosonized picture. It is
further discussed in section 5.

The above state is the only one-particle state of the theory.  All other
massive state are multi particle state which are built from this 'meson'.

The content of massless multi-flavor $QCD_2$
in the large $\nf$ limit is thus very simple and is in fact
closely related to   multi-flavor massless QED\cite{rabin}.
In the latter case model the spectrum  of single particles
contains the following  single state
 \beq
M^2\  J(-p) \state{0} = {e^2 \nf \over \pi} \  J(-p) \state{0}
\eeq
The reason of this similarity  is very clear.
 Neglecting the $O(\nf^{-{1\over 2}})$
 term
in eqn.\eqref{Slnf} corresponds to   dropping the structure constant term in
the
 Affine-Lie  algebra.
In other words, in this limit we perform an  Abelianization of the
theory.

However, one may identify a difference
in the substructure of the  two mesons. The Abelian
meson is built up from two quarks since  current  quanta is made out of  two
quarks,  whereas the non Abelian meson is a more complex object. The difference
can be seen by writing the state explicitly in terms of quarks operators. The
(traceless) current is written as:
\beq
 J^i_j(-p) = \int ^\infty _{-\infty} dq \psi ^{\dagger i}_c(q)\psi ^c_j(-p-q)
 - {{\delta ^i _j} \over N_c}
  \int ^\infty _{-\infty} dq \psi ^{\dagger k}_c(q)\psi ^c_k(-p-q)
\eeq
The flavored adjoint vacuum is written as:
\beq
\state{0} ^{ja}_{ib} = \psi ^{\dagger j}_b(0) \psi ^a_i(0) \state{0}
\eeq
It is a state of two quarks with zero (light-cone) momentum acting on the
identity vacuum.
Thus the massive meson is
\ber
& J^i_j(-p) \state{0}^{ja}_{ib} = (\int ^\infty _{-\infty} dq [\psi ^{\dagger
i}_c(q)\psi ^c_j(-p-q)
 - {{\delta ^i _j} \over N}
 \psi ^{\dagger k}_c(q)\psi ^c_k(-p-q)])\times
  \psi ^{\dagger j}_b(0) \psi ^a_i(0) \state{0}\CR
& =  \int ^p_0 dq \psi ^{\dagger i}_c(-q) \psi ^c_j(-p+q) \psi ^{\dagger
j}_b(0)
 \psi^a_i(0) \state{0}
 - {1\over N} \int ^p_0 dq \psi ^{\dagger i}_c(-q) \psi ^c_i(-p+q)
 \psi ^{\dagger j}_b(0) \psi ^a_j(0) \state{0} \CR
 & + {{N^2-1} \over N} \psi ^{\dagger i}_b(-p) \psi ^a _i(0) \state{0},
\eer
namely,  a color singlet which  is a mixture of four quarks
and two quarks.

The basic feature used in this section has been the  fact that
 $k\gg N_c $. This holds, in fact, not only for large number of fundamental
representations but obviously also for any large level gauged WZW $SU(N)$
model.

\section{summary}
 In this work we have calculated the mesonic
spectra of several \Q  models by
employing bosonization, light-front quantization and expansion in large
number of colors or flavors.

 The main results  of the work are

a)  An approximated spectrum of the ``adjoint fermions" model.

b)  The exact  leading order in  ${1\over \nf}$
 spectra of  multi-flavor fundamental representation.

As for (a),  our approximation is similar to the one used
  in the fermionic picture \cite{kutasov}, namely, dropping the
terms that mix wave-functions with different
number of current creation operators. The physical meaning
 of such approximation is suppressing   pair creation and pair annihilation.
 This  approximation is not quite justified, but it gives us hint about
 the structure of the  spectrum. Obviously, the most urgent task in this
direction  is to look for methods to solve the full wave equation.

The second result states that the spectrum of the large $N_f$ fundamental
 fermions is built out of  a single massive
particle with $M^2\sim e^2 \nf $. For comparison see ref.\cite{Engel} where
 an analysis of similar cases is discussed in the Hamiltonian formalism.
In fact, it is a universal behavior of any gauged $SU(N)$ WZW model with
$k \gg N$.

The interesting question that naturally arise is what happens in the
intermediate region, where $k \sim N$. This region can be realized for
instance,  in the case of
multi-flavor $QCD_2$ with $N_f \sim N_c$.  It is reasonable to expect
 that the  mesonic spectrum  in this region will
lie between that of a single massive state and that with an exponential
density growth. In the absence of exact analytical methods one may have
to invoke
 numerical diagonalization
of the bosonized $M^2$ operator in a similar way to that of the fermionic
picture\cite{gyan}.

For the analysis of the baryonic spectrum in the bosonization approach
it is essential to consider the case of massive quarks\cite{DFS}.
 This maybe   also  the case for the mesonic spectrum since  the mass term
couples the colored, flavored and baryon number sectors.
In fact, even for the massless case a better strategy is to
solve the massive case
and then go to the massless limit.   The extraction of
the mesonic spectrum in the massive case is much more evolved since the
mass term can not be written in a simple fashion in terms of the currents.
However, one can systematically expand the mass term in powers of
${m_q\over e}$ where $m_q$ is the quark mass. Solving the wave equations
in the presence of these massive perturbation deserves a further future
study.

Recently, the theories of   $YM_2$ and  \Q
were analyzed as ``perturbed" topological coset models\cite{FHS}. The spectrum
in that approach  which was deduced using a BRST procedure includes a
peculiar massive state which was not detected in other approaches including
the present.  This discrepancy maybe related to the different approximation
used in that approach.  Clearing up this point as well as the implementation
of that method
  to the case of adjoint fermions and other possible
generalization deserves a further investigation. Other methods inherited
from string theory and conformal field theory can be also be applied to the
analysis of the models discussed\cite{abdalla}.

\Q models can be generalized to a much richer class of theories which are
also gauge invariant and renormalizable\cite{Doug}.
The framework of this  generalization
is the formulation of  the $YM_2$
  functional integral  in terms of
an action which is linear in $F$ and includes  an additional
auxiliary pseudoscalalr field. Ordinary \Q has a quadratic term in the
latter field while taking any arbitrary function  $f$ of this
auxiliary field spans the space of
generalized models. The analysis presented in the present paper
can be applied also to those models.
The  momentum operator $P^-$  rather then being quadratic in
 $ {1\over{\partial _- }}J $ it will take the general form of
$ f({1\over{\partial _- }}J) $.
It will be interesting to compare the outcome of the methods used in the
 current work to those derived in ref.\cite{Doug}.

\Acknowledgements
  We thank  for  M. Engelhardt, Y. Frishman and D. Kutasov for
stimulating conversations. We would specially like to thank S. Yankielowicz
 for numerous discussions in various part of the work.

\section{Appendix}
A detailed calculation of currents commutators.
We would like to calculate $[P^-,J^{j_1}_{j_2}(-p_1)]$,
where $P^-={e^2 \over \pi}\int ^\infty _0 {dp\over {p^2}} J^i_j(-p)J^j_i(p)$.
Therefore we would calculate $[\int ^\infty _0 {dp\over {p^2}} J^i_j(-p)
J^j_i(p),J^{j_1}_{j_2}(-p_1)]$ by using the affine Lie algebra\eqref{Kac}:
\ber
\lefteqn{[\int ^\infty _0 {dp\over {p^2}} J^i_j(-p)J^j_i(p),
J^{j_1}_{j_2}(-p_1)]=} \\ &&
\\ &&
\int ^\infty _0 {dp\over {p^2}} J^i_j(-p)[J^j_i(p),J^{j_1}_{j_2}(-p_1)] +
\int ^\infty _0 {dp\over {p^2}} [J^i_j(-p),J^{j_1}_{j_2}(-p_1)]J^j_i(p) = \\ &&
\\ &&
\int ^\infty _0 {dp\over {p^2}} J^i_j(-p)\{ p{k\over 2}
(\delta ^j_{j_2} \delta ^{j_1}_i
-{1\over N}\delta ^j_i \delta ^{j_1}_{j_2})\delta (p-p_1)  \\ &&
+\half
(J^{j_1}_i(p-p_1)\delta ^j_{j_2}-J^j_{j_2}(p-p_1)\delta ^{j_1}_i) \} \\ &&
+\half \int ^\infty _0 {dp\over {p^2}} (J^{j_1}_j(-p-p_1)\delta ^i_{j_2}
   -J^i_{j_2}(-p-p_1)\delta ^{j_1}_j)J^j_i(p) = \\ &&
\\ &&
{k\over 2p_1}J^{j_1}_{j_2}(-p_1) \\ &&
+\half
\int ^\infty _{-p_1} {dp \over {{(p+p_1)}^2}} J^i_{j_2}(-p-p_1)J^{j_1}_i(p)
-\half
\int ^\infty _{-p_1} {dp \over {{(p+p_1)}^2}} J^{j_1}_j(-p-p_1)J^j_{j_2}(p) \\
&&
+\half
\int ^\infty _0 {dp \over {p^2}} J^{j_1}_j(-p-p_1)J^j_{j_2}(p)
-\half
\int ^\infty _0 {dp \over {p^2}} J^i_{j_2}(-p-p_1)J^{j_1}_i(p) = \\ &&
\eer
The above expression includes annihilation currents as well creation ones.
Separating them one from the other we obtain:
\ber
\lefteqn{={k\over 2p_1}J^{j_1}_{j_2}(-p_1)} \\ &&
+\half \int ^\infty _0 dp ({1 \over {{(p+p_1)}^2}} - {1\over {p^2}})
                                            J^i_{j_2}(-p-p_1)J^{j_1}_i(p) \\ &&
-\half \int ^\infty _0 dp ({1 \over {{(p+p_1)}^2}} - {1\over {p^2}})
                                            J^{j_1}_j(-p-p_1)J^j_{j_2}(p) \\ &&
+\half
\int ^0 _{-p_1} {dp \over {{(p+p_1)}^2}} J^i_{j_2}(-p-p_1)J^{j_1}_i(p) \\ &&
-\half
\int ^0 _{-p_1} {dp \over {{(p+p_1)}^2}} J^{j_1}_j(-p-p_1)J^j_{j_2}(p) = \\ &&
\\ &&
{k\over 2p_1}J^{j_1}_{j_2}(-p_1) \\ &&
+\half NJ^{j_1}_{j_2}(-p_1) \int ^0 _{-p_1} {dp \over {{(p+p_1)}^2}} \\  &&
+\half \int ^0 _{-p_1} dp ({1 \over {{(p+p_1)}^2}} - {1\over {p^2}})
           J^{j_1}_j(p)J^j_{j_2}(-p-p_1) \\ &&
+\half \int ^\infty _0 dp ({1 \over {{(p+p_1)}^2}} - {1\over {p^2}})
  \{ J^i_{j_2}(-p-p_1)J^{j_1}_i(p) - J^{j_1}_i(-p-p_1)J^i_{j_2}(p)\} \\ &&
\eer
In the above expression there are four terms:

The first one ${k\over 2p}J(-p)$,
does not change the number of currents (it has only one current) and it
is proportional to the level $k$. It will play a central role in the large
$k$ limit.

The second term is similar to the first one, but it is proportional to $N$
and it diverges. The divergent part will be compensated by another divergent
term (which arises from the fourth term).

The third term include two creation currents. Thus the interaction $P^-$,
with the help of the algebra created a current. In our discussion we
will ignore this part, for the sake of simplicity.

The last term includes annihilation currents, and therefore we should evaluate
its commutator with other creation current.
\ber
\lefteqn{[\half \int ^\infty _0 dp ({1\over {{(p+p_1)}^2}} - {1\over {p^2}})
  \{ J^i_{j_2}(-p-p_1)J^{j_1}_i(p) - J^{j_1}_i(-p-p_1)J^i_{j_2}(p)\},
   J^{j_2}_{j_3}(-p_2)]=} \\ &&
\\ &&
\half \int ^\infty _0 dp ({1\over {{(p+p_1)}^2}} - {1\over {p^2}})
\{ J^i_{j_2}(-p-p_1)[J^{j_1}_i(p),J^{j_2}_{j_3}(-p_2)] \\ &&
+[J^i_{j_2}(-p-p_1),J^{j_2}_{j_3}(-p_2)]J^{j_1}_i(p)
-J^{j_1}_i(-p-p_1)[J^i_{j_2}(p),J^{j_2}_{j_3}(-p_2)] \\ &&
-[J^{j_1}_i(-p-p_1),J^{j_2}_{j_3}(-p_2)]J^i_{j_2}(p)\} = \\ &&
\\ &&
\half \int ^\infty _0 dp ({1\over {{(p+p_1)}^2}} - {1\over {p^2}})
\{ J^i_{j_2}(-p-p_1)p{k\over 2}(\delta ^{j_2}_i \delta ^{j_1}_{j_3} - {1\over
N}
\delta ^{j_1}_i \delta ^{j_2}_{j_3}) \delta (p-p_2) \\ &&
\half J^i_{j_2}(-p-p_1)J^{j_2}_i(p-p_2)\delta ^{j_1}_{j_3}
-\half J^i_{j_2}(-p-p_1)J^{j_1}_{j_3}(p-p_2) \delta ^{j_2}_i \\ &&
-\half NJ^i_{j_3}(-p-p_1-p_2)J^{j_1}_i(p) \\ &&
-J^{j_1}_i(-p-p_1)p{k\over 2}(\delta ^i_{j_3} \delta ^{j_2}_{j_2} - {1\over N}
\delta ^i_{j_2} \delta ^{j_2}_{j_3}) \delta (p-p_2) \\ &&
+\half NJ^{j_1}_i(-p-p_1)J^i_{j_3}(p-p_2) \\ &&
-\half J^{j_2}_i(-p-p_1-p_2)J^i_{j_2}(p)\delta ^{j_1}_{j_3}
+\half J^{j_1}_{j_3}(-p-p_1-p_2)J^i_{j_2}(p)\delta ^{j_2}_i \} =
\eer
This leads to the following expression:
\ber
\lefteqn {=-\half p_2 N{k\over 2}({1\over {{(p_1+p_2)}^2}} - {1\over
{{p_2}^2}})
J^{j_1}_{j_3}(-p_2-p_1)} \\ &&
+{1\over 2}
\int ^\infty _0 dp ({1\over {{(p+p_1)}^2}} - {1\over {p^2}}) \times \\ &&
\times \{ J^i_j(-p-p_1)J^j_i(p-p_2)-J^j_i(-p-p_1-p_2)J^i_j(p)\}
 \delta ^{j_1}_{j_3} \\ &&
 +{N\over 2}
 \int ^\infty _0 dp ({1\over {{(p+p_1)}^2}} - {1\over {p^2}}) \times \\ &&
 \times
 \{ J^{j_1}_i(-p-p_1)J^i_{j_3}(p-p_2)-J^i_{j_3}(-p-p_1-p_2)J^{j_1}_i(p)\}
\eer
A few remarks about the last expression:

The first term includes only creation currents, therefore there is no need to
evaluate its commutators with other currents.

The second term contains creation and annihilation currents, but it is
suppressed in the large $N$ limit.

The last term is an important term. It is not suppressed at large $N$, and
it contains the following expression in it:
\beq
{N\over 2} \int ^{p_2} _0 dp ({1\over {{(p+p_1)}^2}} - {1\over {p^2}})
J^{j_1}_i(-p-p_1)J^i_{j_3}(p-p_2)
\eeq
Which may rewritten as:
\beq
{N\over 2}
\int ^{p_2} _0 dp ({1\over {{(p_1+p_2-p)}^2}} - {1\over {{(p_2-p)}^2}})
J^{j_1}_i(p-p_1-p_2)J^i_{j_3}(-p)
\eeq
This expression leads to the generalized 't Hooft equation.

\end{document}